\documentclass[twocolumn,superscriptaddress,amsmath,amssymb,aps,prl]{revtex4-2}

\usepackage{graphicx}
\usepackage{dcolumn}
\usepackage{ulem}
\usepackage{bm}
\usepackage{amsthm}
\newtheorem*{theorem*}{Theorem}

\usepackage[colorlinks,urlcolor=blue,citecolor=blue,linkcolor=blue]{hyperref}

\begin{document}
	\title{
    Quasi-periodic moir\'{e} patterns and dimensional localization in three-dimensional quasi-moir\'{e} crystals
    }

	\author{Ce Wang}
	\affiliation{School of Physics Science and Engineering, Tongji University, Shanghai, 200092, China}
	\author{Chao Gao}
	\email{gaochao@zjnu.edu.cn}
	\affiliation{Department of Physics, Zhejiang Normal University, Jinhua, 321004, China}
    \affiliation{Zhejiang Institute of Photoelectronics, Jinhua 321004, China}

	\author{Zhe-Yu Shi}
	\email{zyshi@lps.ecnu.edu.cn}
	\affiliation{State Key Laboratory of Precision Spectroscopy, East China Normal University, Shanghai 200062, China}
	\date{\today}
	
	\begin{abstract}
        Recent advances in spin-dependent optical lattices [Meng et al., Nature \textbf{615}, 231 (2023)] have enabled the experimental implementation of two superimposed three-dimensional lattices, presenting new opportunities to investigate \textit{three-dimensional moir\'{e} physics} in ultracold atomic gases. This work studies the moir\'{e} physics of atoms within a spin-dependent cubic lattice with relative twists along different directions. It is discovered that dimensionality significantly influences the low-energy moir\'{e} physics. From a geometric perspective, this manifests in the observation that moir\'{e} patterns, generated by rotating lattices along different axes, can exhibit either periodic or quasi-periodic behavior—a feature not present in two-dimensional systems. We develop a low-energy effective theory applicable to systems with arbitrary rotation axes and small rotation angles. This theory elucidates the emergence of quasi-periodicity in three dimensions and demonstrates its correlation with the arithmetic properties of the rotation axes. Numerical analyses reveal that these quasi-periodic moir\'{e} potentials can lead to distinctive dimensional localization behaviors of atoms,  manifesting as localized wave functions in planar or linear configurations.
	\end{abstract}

	\maketitle

When two identical two-dimensional (2d) lattices are overlaid with a slight twist, periodic interference patterns known as moir\'{e} patterns emerge at the macroscopic scale. These patterns have garnered significant attention, as recent studies of twisted double-layered 2d materials demonstrate that moir\'{e} patterns can significantly alter systems' electronic properties. Remarkable phenomena have been discovered in 2d moir\'{e} systems, including flat bands~\cite{bistritzer2011moire,cao2018correlated,tarnopolsky2019origin}, unconventional superconductivity~\cite{cao2018unconventional,wu2018theory,isobe2018unconventional,lian2019twisted}, and fractional quantum anomalous Hall effect~\cite{cai23nature,park23nature,xu2023observation,lu_2024}.

While moir\'{e} patterns are predominantly observed in 2d materials, they can be engineered in various synthetic structures, including photonic crystals~\cite{18science,wang2020localization,mao2021magic,Dong21PRL,Lou_2022,Tang_2023}, phononic crystals~\cite{zheng2020phonon,duan2020twisted,hu2020topological,chen2020configurable,Yao24PRL}, lattice nanocavities~\cite{23FR,luan2023reconfigurable,spencer2025van}, and optical lattices~\cite{gonzalez2019cold,20PRL,Luo_2021,Paul23PRA,meng2023atomic,wan2024fractal,zeng2024dynamical,tian2024nonlinearity,fang2025bifurcations}. These structures provide versatile platforms for exploring moir\'{e} physics that are challenging to study in conventional condensed matter systems. Notably, recent experimental advances in synthetic moir\'{e} superlattices have demonstrated interesting moir\'{e} physics for bosonic matters, particularly the localization of electromagnetic waves~\cite{wang2020localization,23PRL} and novel superfluid to Mott insulator transitions of bosonic atoms~\cite{meng2023atomic}. We have also proposed that through an extension of the current twisted bilayer optical lattice configuration, it is possible to implement a twisted three-dimensional (3d) optical lattice for ultracold atomic gases~\cite{Wang24PRL}. This offers a unique opportunity for the study of \textit{3d moir\'{e} physics}, providing insights previously inaccessible through conventional condensed matter materials.

Building upon our previous work~\cite{Wang24PRL}, which establishes the complete classification for 3d moir\'{e} crystals, i.e. lattices with commensurate twists, this research explores incommensurate twist configurations, termed quasi-moir\'{e} crystals. Specifically, we focus on moir\'{e} physics in twisted cubic lattices with small twist angles and probe the role of dimensionality on moir\'{e} physics.

\begin{figure*}[t]
  \includegraphics[width=0.95\textwidth]{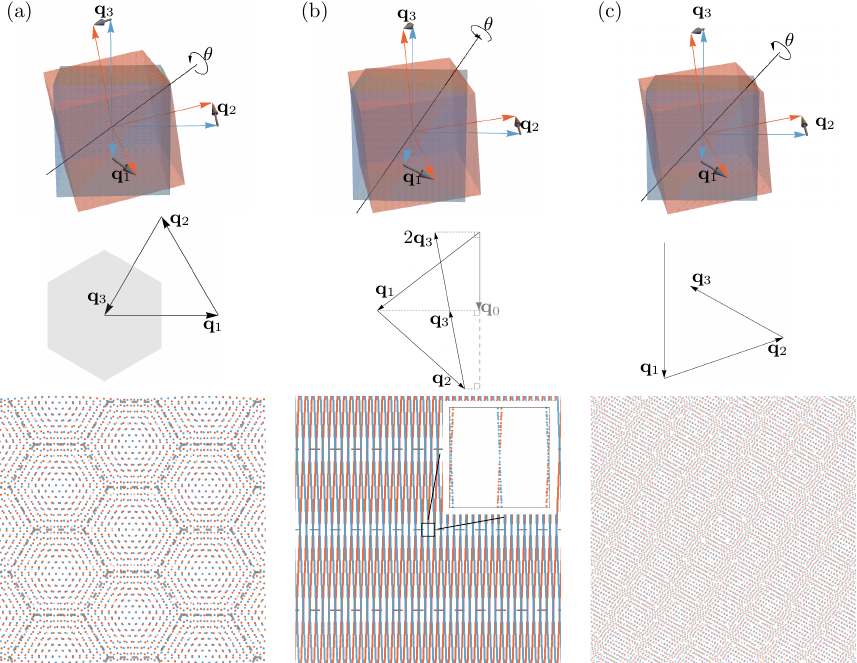}

  \caption{Schematic plot of the moir\'{e} primitive vectors $\mathbf{q}_i$ and the moir\'{e} patterns for twisted cubic lattices with rotation axes $\mathbf{L}=(1,1,1)$ (column (a), periodic type), $\mathbf{L}=(1,\varphi,\varphi^2)$ (column (b), the golden axis, hybrid type), and $\mathbf{L}=(1,\sqrt{\varphi},\varphi)$ (column (c), the Kepler axis, ergodic type). Here $\varphi=\frac{\sqrt{5}+1}{2}$ is the golden ratio. The first row illustrates the moir\'{e} primitive vectors $\mathbf{q}_i$ in 3d momentum space. The blue (red) cubes represent the first Brillouin zones of the underlying A (B) lattices. The middle row illustrates the relations between moir\'{e} primitive vectors by plotting them on the plane perpendicular to $\mathbf{L}$. The gray area in (a) and the line segment in (b) mark the period of $S_\mathbf{q}\equiv\{n_1\mathbf{q}_1+n_2\mathbf{q}_2+n_3\mathbf{q}_3\}$. The last row displays the corresponding moir\'{e} patterns. Each diagram is generated by projecting $2\times41^3$ lattice points onto the plane perpendicular to the rotation axis $\mathbf{L}$. The dashed lines mark the periodicity calculated using the moir\'{e} primitive vectors $\mathbf{q}_i$. One can see that the moir\'{e} pattern of a quasi-moir\'{e} crystal can be either 2d periodic (panel (a)), quasi-periodic in all directions (panel (c)), or periodic in one direction while quasi-periodic in others (panel (b)).}  \label{fig1}
\end{figure*}

Before entering into any detailed analysis, we shall first provide some geometric intuitions that highlight the significance of dimensionality. The key differentiation between moir\'{e} physics in two and three dimensions stems from the distinctive parameterization of rotations in two and three dimensions. In 2d systems, the rotation is characterized by a single parameter, i.e., the twist angle $\theta$. Consequently, the structure of the moir\'{e} pattern is uniquely determined by the underlying lattice, resulting in \textit{periodic} patterns. Adjusting the twist angle simply scales the whole pattern by a dimensionless factor. In contrast, a 3d rotation is determined by both the twist angle $\theta\in[0,\pi)$ and the rotation axis $\mathbf{L}\in\mathbb{R}^3$. The parameterization enables the generation of more diverse moir\'{e} patterns from a single 3d lattice. In Fig.~\ref{fig1}, we plot the moir\'{e} patterns of twisted cubic lattices with various rotation axes $\mathbf{L}$ by projecting lattice points onto 2d planes perpendicular to it. The resulting patterns exhibit remarkable diversity, manifesting as periodic (Fig.~\ref{fig1}a), quasi-periodic (Fig.~\ref{fig1}c), or hybrid structures that display periodicity along one direction while maintaining quasi-periodicity along the others (Fig.~\ref{fig1}b).

This work addresses two central questions regarding 3d quasi-moir\'{e} crystals: What mechanisms drive these diverse moir\'{e} patterns, and whether these quasi-periodic patterns lead to wave function localizations? To explore these queries, we formulate a low-energy effective theory for small-angle rotations about a fixed axis, building upon Bistritzer and Macdonald's work~\cite{bistritzer2011moire}. Our theoretical framework elucidates the emergence of quasi-periodic moiré patterns and provides a method for classifying different types of moir\'{e} patterns based on the arithmetic property of $\mathbf{L}$. Through numerical analysis, we demonstrate the potential for distinct dimensional localizations within quasi-moir\'{e} crystals with different rotation axes.

\textit{Model }---
We consider the following Hamiltonian that describes cold atoms moving in a spin-dependent lattice,
\begin{align}
	H=\left(
\begin{matrix}
 \frac{\mathbf{p}^2}{2m_0}+V_A& \Omega \\
 \Omega & \frac{\mathbf{p}^2}{2m_0}+V_B
\end{matrix}
\right),\label{Hamiltonian}
\end{align}
where the two components represent the two hyperfine states (spins) of an atom with mass $m_0$ and momentum $\mathbf{p}$. The atoms in two states experience spin-dependent lattices $V_A$ and $V_B$, and are coupled by an external radio-frequency field with coupling strength $\Omega$. 

We consider the case that both $V_A$ and $V_B$ are identical cubic lattices with a relative twist $\mathbf{R}\in\text{SO}(3)$, i.e., $V_A(\mathbf{r})=V(R^{\frac{1}{2}}\mathbf{r})$ and $V_B(\mathbf{r})=V(R^{-\frac{1}{2}}\mathbf{r})$ with $V(\mathbf{r})={V_0}\left[\sin^2(\pi x)+\sin^2(\pi y)+\sin^2(\pi z)\right]$. Here, $V_0$ represents the potential depth, $R=R(\mathbf{L},\theta)$ is the rotation matrix corresponding to axis $\mathbf{L}$ and angle $\theta$, and $R^{\pm \frac{1}{2}}=R(\mathbf{L,\pm\theta/2})$. We also assume the recoil momentum $k_r=\pi$ for simplicity. We note that the square lattice version of Hamiltonian~\eqref{Hamiltonian} has been recently realized in ultracold ${}^{87}$Rb gas by the Shanxi University group~\cite{meng2023atomic}. Our model is a direct generalization of their current experimental setup.

\textit{Low-energy effective theory }---
In the previous work~\cite{Wang24PRL}, we proved that the twisted lattice forms a periodic moir\'{e} crystal if and only if one can find coprime integers $l_1,l_2,l_3$ such that $\mathbf{L}=(l_1,l_2,l_3)$ and the twist angle $\theta$ satisfies
$\theta=\arccos\frac{m^2-n^2 \mathbf{L}^2}{m^2+n^2 \mathbf{L}^2}$, for some integer $m,n\in\mathbb{Z}$. Note that all the commensurate rotations represent only a measure zero subset of $\text{SO}(3)$, due to constraints on both rotation axis and twist angle. The incommensurate quasi-moir\'{e} crystals hence form a significantly broader parameter space compared to moir\'{e} crystals. The absence of periodicity also leads to additional complexities in numerical investigations of quasi-moir\'{e} crystals. It is thus useful to develop an effective theory that describes the system's low-energy behavior.

We rewrite the Hamiltonian as $H=H_{A}+H_{B}+H_{AB}$, where $H_{A}$ ($H_B$) is the diagonal part that describes atoms moving in lattice $V_A$ ($V_B$), and $H_{AB}$ represents the off-diagonal term that couples the two spin components. In momentum space, the diagonal Hamiltonians can be written as ($\hbar=1$, only the lowest band is considered) $H_\sigma=\sum_{\mathbf{p}}\epsilon_{\sigma}(\mathbf{p})|\mathbf{p},\sigma\rangle\langle\mathbf{p},\sigma|$. Here, $|\mathbf{p},\sigma\rangle$ ($\sigma=A,B$) represents the momentum $\mathbf{p}$ state with corresponding spin, $\epsilon_A(\mathbf{p})=\epsilon(R^{\frac{1}{2}}\mathbf{p})$ ($\epsilon_B(\mathbf{p})=\epsilon(R^{-\frac{1}{2}}\mathbf{p})$) is the dispersion for the A (B) lattice. In the tight-binding limit ($V_0\gg E_r$ with $E_r=\frac{\hbar^2\pi^2}{2m_0}$ being the recoil energy), one has $\epsilon(\mathbf{p})=-2J(\cos p_x+\cos p_y+\cos p_z)$ with $J$ being the tunneling coefficient between nearest sites~\cite{SM}.

We focus on low-energy states near the $\Gamma$ point, i.e. $\mathbf{p}\simeq0$. In this region, the effective-mass expansion gives
\begin{align}
\epsilon_{A}(\mathbf{p})\simeq\epsilon_{B}(\mathbf{p})\simeq\frac{\mathbf{p}^2}{2m_{\text{eff}}}-6J\label{diagonal}
\end{align}
with the effective mass $m_\text{eff}=(2J)^{-1}$.
The coupling between the two spin states is
\begin{align}
&\quad\langle \mathbf{p},A|H_{AB}|\mathbf{p}',B\rangle\nonumber\\
&=8\pi^{3/2}\Omega l_{\text{ho}}^3\sum_{\mathbf{G}_A,\mathbf{G}_B}e^{-(\mathbf{p}+\mathbf{G}_A)^2l_{\text{ho}}^2}\delta_{\mathbf{p}+\mathbf{G}_A,\mathbf{p}'+\mathbf{G}_B},\label{coupling}
\end{align}
where $l_{\text{ho}}=(4m_0^2V_0E_r)^{-1/4}\ll1$ is the width of the Wannier function, $\mathbf{G}_A\in2\pi R^{-\frac{1}{2}}\mathbb{Z}^3$ ($\mathbf{G}_B\in2\pi R^{\frac{1}{2}}\mathbb{Z}^3$) represents the reciprocal momentum of the A (B) lattice.

Note that the coupling strength decays exponentially when $|\mathbf{G}_A|\gtrsim l_{\text{ho}}^{-1}$. Hence, in the spirit of Bistritzer and Macdonald~\cite{bistritzer2011moire}, one concludes that when the twist angle is small such that $\theta\cdot l_\text{ho}^{-1}\ll1$, the $\mathbf{G}_B$ terms that contribute to the summation of Eq.~\eqref{coupling} are those satisfy the condition $\mathbf{G}_B=R\mathbf{G}_A$.
Under the above approximation, we can Fourier transform both Eq.~\eqref{diagonal} and \eqref{coupling} to the real-space representation, which leads to an effective Hamiltonian
\begin{align}
H_{\text{eff}}=\frac{\mathbf{p}^2}{2m_\text{eff}}+V_m(\mathbf{r})\sigma_x
\label{Heff}
\end{align}
with
\begin{align}
V_m(\mathbf{r})=8\pi^{3/2}\Omega l_{\text{ho}}^3\sum_{\mathbf{G}\in2\pi\mathbb{Z}^3}e^{-\mathbf{G}^2l_{\text{ho}}^2}e^{i(R^{\frac{1}{2}}-R^{-\frac{1}{2}})\mathbf{G}\cdot\mathbf{r}}.\label{Vm}
\end{align}
Utilizing the symmetry between the two spin states, we can further diagonalize $H_\text{eff}$ by considering an even or odd combination of two spin states. This gives a single component effective Hamiltonian $H_{\text{eff},\pm}=\frac{\mathbf{p}^2}{2m_\text{eff}}\pm V_m(\mathbf{r})$, where $V_m(\mathbf{r})$ can be viewed as an effective potential generated by the moir\'{e} pattern.

\textit{Emergence of quasi-periodic moir\'{e} patterns }---
The structure and periodicity of a moir\'{e} pattern are closed related to the moir\'{e} potential $V_m$, as both of them reflect the variation in local alignment (overlap) between two lattices across an extended spatial scale (in the order of $\theta^{-1}$). Through detailed analysis of Eq.~\eqref{Vm}, we can identify several key characteristics of the moir\'{e} potential that illuminate the properties of the corresponding moir\'{e} patterns.

First, by separating the position vector $\mathbf{r}$ into components parallel and perpendicular to the rotation axis as $\mathbf{r}=\mathbf{r}_\parallel+\mathbf{r}_\perp$, it can be checked that the moir\'{e} potential $V_m$ only depends on the perpendicular component $\mathbf{r}_\perp$. This observation demonstrates that the moir\'{e} pattern of a twisted 3d lattice manifests in two dimensions, thereby explaining our ability to visualize the moir\'{e} pattern in Fig.~\ref{fig1} through projecting 3d lattice points onto a plane perpendicular to $\mathbf{L}$. The property also establishes the translational invariance of $H_{\text{eff}}$ along the axis $\mathbf{L}$, enabling the reduction of the Schr\"{o}dinger equation to two dimensions and simplifying the numerical calculations.

Furthermore, when considering a given rotation axis $\mathbf{L}$, the moir\'{e} potentials corresponding to different twist angles $\theta_1$ and $\theta_2$ exhibit a scaling relation,
\begin{align}
V_{m,1}(\lambda_{\theta_1}^{-1}\mathbf{r})=V_{m,2}(\lambda_{\theta_2}^{-1}\mathbf{r}),
\end{align}
where $\lambda_\theta=2\sin\theta/2\simeq\theta$ represents a dimensionless scaling parameter. The relation demonstrates that varying the twist angle $\theta$ results solely in a spatial scaling, while maintaining the fundamental structure of the moir\'{e} pattern.

\begin{figure}[t]
  \includegraphics[width=0.5\textwidth]{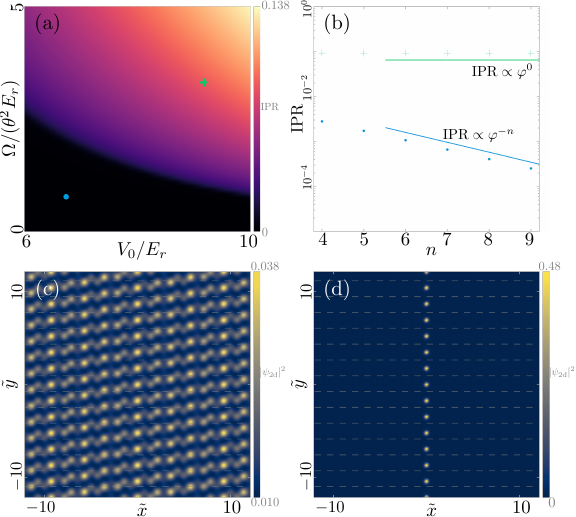}

  \caption{Dimensional localization for quasi-moir\'{e} crystal twisted around the golden axis $\mathbf{L}=(1,\varphi,\varphi^2)$ (hybrid type). (a) The IPR as a function of potential depth $V_0/E_r$ and normalized coupling strength $\lambda_\theta^{-2}\Omega/E_r\simeq\Omega/(\theta^2E_r)$. (b) The scaling of the IPR as a function of the order of the rational approximation of $\mathbf{L}=(1,\varphi,\varphi^2)\approx(F_n,F_{n+1},F_{n+2})$, where $F_n$ is the $n$-th Fibonacci number. (c,d) The 2d wave function in the plane perpendicular to $\mathbf{L}$. The parameters correspond to the blue (panel (c)) and green (panel (d)) dots in panel (a) respectively. $\tilde{x}=\lambda_\theta x$ and $\tilde{y}=\lambda_\theta y$ are the rescaled coordinates with $y$ represents the direction of $-\mathbf{q}_0$. The dashed lines mark the periodicity calculated using $\mathbf{q}_0$. 
}  \label{fig2}
\end{figure}

\begin{figure}[t]
  \includegraphics[width=0.5\textwidth]{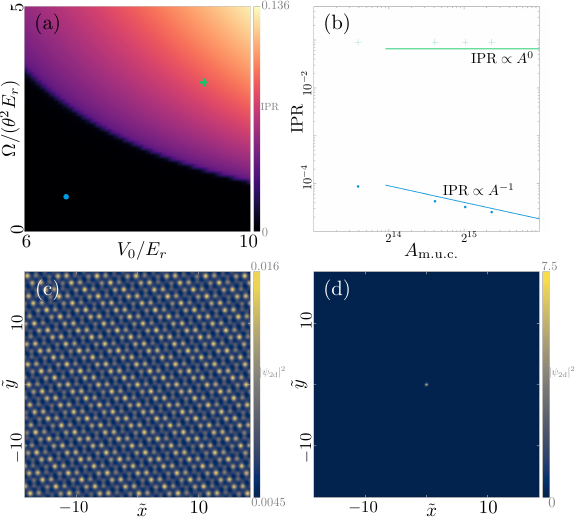}

  \caption{Dimensional localization for quasi-moir\'{e} crystal twisted around the Kepler axis $\mathbf{L}=(1,\sqrt{\varphi},\varphi)$ (ergodic type). (a) The IPR as a function of potential depth $V_0/E_r$ and normalized coupling strength $\lambda_\theta^{-2}\Omega/E_r\simeq\Omega/(\theta^2E_r)$. (b) The scaling of the IPR as a function of $A_{\text{m.u.c.}}$, i.e., the area of the moir\'{e} unit cell under rational approximation. (c,d) The 2d wave function in the plane perpendicular to $\mathbf{L}$. The parameters correspond to the blue (panel (c)) and green (panel (d)) dots in panel (a) respectively. $\tilde{x}=\lambda_\theta x$ and $\tilde{y}=\lambda_\theta y$ are the rescaled coordinates with $y$ represents the direction of $-\mathbf{q}_1$.}  \label{fig3}
\end{figure}

Third, all the momentum transfers induced by the moir\'{e} potential follow the form $(R^{\frac{1}{2}}-R^{-\frac{1}{2}})\mathbf{G}$. While the set of $\mathbf{G}$ vectors is spanned by $\mathbf{G}_1=2\pi(1,0,0)^\text{T}$, $\mathbf{G}_2=2\pi(0,1,0)^\text{T}$, and $\mathbf{G}_3=2\pi(0,0,1)^\text{T}$. The momentum transfers of $V_m$ is thus spanned by
\begin{align}
\mathbf{q}_i=(R^{\frac{1}{2}}-R^{-\frac{1}{2}})\mathbf{G}_i=\lambda_\theta\frac{\mathbf{L}}{|\mathbf{L}|}\times\mathbf{G}_i,\ i=1,2,3,\label{qi}
\end{align}
which are illustrated in Fig.~\ref{fig1}.

This observation reveals a fundamental distinction between the moir\'{e} patterns in 2d and 3d lattices, i.e., specifically in terms of the number of $\mathbf{q}_i$ vectors. In 2d lattices, the set of transfer momenta is generated by only two $\mathbf{q}_i$ vectors because $\mathbf{G}_i$ are the basis of a 2d reciprocal lattice. As a result, the moir\'{e} potentials (patterns) must exhibit periodicity, with their structure directly corresponding to the underlying lattice periods. This occurs because $\mathbf{q}_i\propto \mathbf{G}_i$, while the remaining coefficient and cross multiplication in Eq.~\eqref{qi} merely introduces a scaling and a $\pi/2$ rotation.

In contrast, 3d quasi-moir\'{e} crystals present three $\mathbf{q}_i$ vectors lying in the plane perpendicular to $\mathbf{L}$. This configuration presents a distinctly different scenario, as they span a set $S_{\mathbf{q}}\equiv\{n_1\mathbf{q}_1+n_2\mathbf{q}_2+n_3\mathbf{q}_3|n_i\in\mathbb{Z}\}$ that does not necessarily constitute a 2d Bravais (reiprocal) lattice. In the supplementary material, we prove that all the moir\'{e} patterns can be categorized into three types, based on the arithmetic properties of the rotation axis $\mathbf{L}=(l_1,l_2,l_3)$:

I. The \textit{periodic} type: When both $l_2/l_1$ and $l_3/l_1$ are rationals, there exists a non-trivial integral combination of $\mathbf{q}_i$ where $\sum_i n_i\mathbf{q}_i=0$.  In this instance, the set $S{\mathbf{q}}$ forms a 2d Bravais lattice, resulting in a periodic moir\'{e} pattern. This is exemplified by $\mathbf{L}=(1,1,1)$, as shown in Fig.~\ref{fig1}(a).

II. The \textit{ergodic} type: When $l_2/l_1$ and $l_3/l_1$ are two linearly independent irrational numbers on the rational field~\cite{linear_independence}, the set $S_{\mathbf{q}}$ becomes dense on the 2d momentum plane. The moir\'{e} pattern in this instance exhibits quasi-periodicity in all planar directions. This is exemplified by $\mathbf{L}=(1,\sqrt{\varphi},\varphi)$, where $\varphi=\frac{\sqrt{5}+1}{2}$ being the golden ratio, as shown in Fig.~\ref{fig1}(c).

III. The \text{hybrid} type: When at least one of $l_2/l_1$ and $l_3/l_1$ is irrational, and they are linearly dependent on the rational field, there exists a finite vector $\mathbf{q}_0$ such that the set $S_{\mathbf{q}}$ becomes dense on infinite parallel lines separated by $\mathbf{q}_0$, with these lines oriented perpendicular to $\mathbf{q}_0$. In this instance, the moir\'{e} pattern is periodic along $\mathbf{q}_0$ with a period $2\pi/|\mathbf{q}_0|$ while maintaining quasi-periodicity in other directions. This case is exemplified by $\mathbf{L}=(1,\varphi,\varphi^2)$, as shown in Fig.~\ref{fig1}(b), where $\mathbf{q}_0=(\mathbf{q}_1+\mathbf{q}_2-\mathbf{q}_3)/3$.


\textit{Dimensional localization}---
It is well known that quantum particles moving in a quasi-periodic potential might lead to wave function localization. While our understanding of quasi-periodic systems has primarily focused on one-dimensional models, the field of higher-dimensional quasi-periodic models remains relatively unexplored. The ergodic and the hybrid quasi-moir\'{e} crystals thus present a valuable opportunity to advance both experimental and theoretical research of novel localization transitions in two dimensions. To examine the localization characteristics of both types of moir\'{e} patterns, we conduct numerical analysis through the inverse participation ratio (IPR) of single-particle ground states.

In examining the hybrid type, we focus our analysis on the rotation axis $\mathbf{L}=(1,\varphi,\varphi^2)$, designated as the golden axis. Our methodology involves transforming the problem into a periodic case by approximating the rotation axis ``rationally'', i.e., $\mathbf{L}\simeq(F_n,F_{n+1},F_{n+1})$ with $F_n$ being the $n$-th Fibonacci number. The approximation enables us to efficiently compute the ground state wave function of the IPR within the approximate moir\'{e} unit cell. The numerical results, presented in Fig.~\ref{fig2}(a), demonstrates a distinct transition between states where $\text{IPR}\simeq0$ and those where $\text{IPR}=O(1)$. Fig.~\ref{fig2}(c) and (d) illustrate the characteristic spatial distributions of wave functions in these regions. Due to the moir\'{e} pattern's periodicity in the $\mathbf{q}_0$ ($\tilde{y}$) direction, the density profile in the
``localized'' region manifests an infinitely long line. Note that the moir\'{e} potential $V_m$ is translationally invariant along the $\mathbf{L}$ direction. Consequently, states within the ``localized'' region are constrained to a 2d plane, exhibiting localization in the $\tilde{x}$ direction while maintaining extension in both the $\tilde{y}$ and $\mathbf{L}$ directions. We have termed this phenomenon a dimensional localized state. This dimensional localization is substantiated through analysis of the scaling of the IPRs with increasing rational approximation order $n$. Given that the approximate moir\'{e} unit cell period scales as $\varphi^n$ in the $\tilde{x}$ direction and $\varphi^0$ in the $\tilde{y}$ direction for large $n$, we anticipate that the IPR of a dimensional localized (extended) state will scale as $\varphi^{-n}$ ($\varphi^0$). This behavior has been numerically validated, as demonstrated in Fig.~\ref{fig2}(b).

In examining the ergodic tye, we analyze the rotation axis $\mathbf{L}=(1,\sqrt{\varphi},\varphi)$, referred to as the Kepler axis due to its components $l_1,l_2,l_3$ forming the sides of a right triangle known as the Kepler triangle (Note $1+\varphi=\varphi^2$)~\cite{Kepler}. Analogous to the hybrid case, we observe a transition between regions with $\text{IPR}\simeq0$ and $\text{IPR}=O(1)$ in Fig.~\ref{fig3}(a). The spatial distribution of the wave function exhibits localization characteristics in all directions (as illustrated in Fig.~\ref{fig3}(c), (d)), attributed to the ergodic property of the moir\'{e} pattern. The transition line depicted in Fig.~\ref{fig3}(a) thus demonstrates a dimensional localization from 3d extended states to confined states on a 1d confined state along the $\mathbf{L}$ direction. This dimensional localization is further validated through numerical analysis of the IPR scaling as a function of the area of the moir\'{e} unit cell ($A_{\text{m.u.c.}}$), as presented in Fig.~\ref{fig3}(b).

It is worth noting that previous studies on 2d twisted lattices have also demonstrated localization transitions~\cite{huang2016localization,Huang19PRB,Moon19PRB,Park19PRB,wang2020localization,tang2021modeling,Gonçalves_2022,23PRL,Gao23PRA,23FR,24PRB,madroñero2024}. While the localization mechanisms in these 2d systems primarily depend on flat moir\'{e} bands or incommensurate lattices with finite twist angles, with both mechanisms exhibiting significant sensitivity to the twist angle $\theta$. The dimensional localization phases we observe in 3d quasi-moir\'{e} crystals, however, demonstrate substantial stability across an extensive parameter region, as evidenced in Fig.~\ref{fig2} and \ref{fig3}.

\textit{Summary} ---
In summary, we examine 3d moir\'{e} physics in the context of ultracold atoms confined in twisted lattices. Our findings establish that dimensionality significantly influences the low-energy physics of quasi-moiré crystals. From a geometric perspective, this influence is evidenced by the emergence of quasi-periodic moiré patterns, a phenomenon absent in 2d systems. Through the development of a low-energy effective theory, we provide a comprehensive explanation for the emergence of quasi-periodicity in the moir\'{e} pattern. Numerical analysis of the effective theory demonstrates that the quasi-periodic moir\'{e} potentials can lead to novel dimensional localization transitions of atomic wave packets.

\begin{acknowledgements}
\textit{Acknowledgement }---
We thank Hui Zhai, Biao Lian, Jing Zhang, and Ya-Xi Shen for inspiring discussion.
This work is supported by the National Natural Science Foundation of China under Grants Nos. 12204352 (C.W.) and 12474493 (C. G.), and the Natural Science Foundation of Zhejiang Province, China under Grants No. LR22A040001 (C. G.).

\end{acknowledgements}

\bibliography{ref}

\begin{thebibliography}{48}%
\makeatletter
\providecommand \@ifxundefined [1]{%
 \@ifx{#1\undefined}
}%
\providecommand \@ifnum [1]{%
 \ifnum #1\expandafter \@firstoftwo
 \else \expandafter \@secondoftwo
 \fi
}%
\providecommand \@ifx [1]{%
 \ifx #1\expandafter \@firstoftwo
 \else \expandafter \@secondoftwo
 \fi
}%
\providecommand \natexlab [1]{#1}%
\providecommand \enquote  [1]{``#1''}%
\providecommand \bibnamefont  [1]{#1}%
\providecommand \bibfnamefont [1]{#1}%
\providecommand \citenamefont [1]{#1}%
\providecommand \href@noop [0]{\@secondoftwo}%
\providecommand \href [0]{\begingroup \@sanitize@url \@href}%
\providecommand \@href[1]{\@@startlink{#1}\@@href}%
\providecommand \@@href[1]{\endgroup#1\@@endlink}%
\providecommand \@sanitize@url [0]{\catcode `\\12\catcode `\$12\catcode `\&12\catcode `\#12\catcode `\^12\catcode `\_12\catcode `\%12\relax}%
\providecommand \@@startlink[1]{}%
\providecommand \@@endlink[0]{}%
\providecommand \url  [0]{\begingroup\@sanitize@url \@url }%
\providecommand \@url [1]{\endgroup\@href {#1}{\urlprefix }}%
\providecommand \urlprefix  [0]{URL }%
\providecommand \Eprint [0]{\href }%
\providecommand \doibase [0]{https://doi.org/}%
\providecommand \selectlanguage [0]{\@gobble}%
\providecommand \bibinfo  [0]{\@secondoftwo}%
\providecommand \bibfield  [0]{\@secondoftwo}%
\providecommand \translation [1]{[#1]}%
\providecommand \BibitemOpen [0]{}%
\providecommand \bibitemStop [0]{}%
\providecommand \bibitemNoStop [0]{.\EOS\space}%
\providecommand \EOS [0]{\spacefactor3000\relax}%
\providecommand \BibitemShut  [1]{\csname bibitem#1\endcsname}%
\let\auto@bib@innerbib\@empty
\bibitem [{\citenamefont {Bistritzer}\ and\ \citenamefont {MacDonald}(2011)}]{bistritzer2011moire}%
  \BibitemOpen
  \bibfield  {author} {\bibinfo {author} {\bibfnamefont {R.}~\bibnamefont {Bistritzer}}\ and\ \bibinfo {author} {\bibfnamefont {A.~H.}\ \bibnamefont {MacDonald}},\ }\bibfield  {title} {\bibinfo {title} {Moir{\'e} bands in twisted double-layer graphene},\ }\href {https://doi.org/10.1073/pnas.1108174108} {\bibfield  {journal} {\bibinfo  {journal} {Proceedings of the National Academy of Sciences}\ }\textbf {\bibinfo {volume} {108}},\ \bibinfo {pages} {12233} (\bibinfo {year} {2011})}\BibitemShut {NoStop}%
\bibitem [{\citenamefont {Cao}\ \emph {et~al.}(2018{\natexlab{a}})\citenamefont {Cao}, \citenamefont {Fatemi}, \citenamefont {Demir}, \citenamefont {Fang}, \citenamefont {Tomarken}, \citenamefont {Luo}, \citenamefont {Sanchez-Yamagishi}, \citenamefont {Watanabe}, \citenamefont {Taniguchi}, \citenamefont {Kaxiras}, \citenamefont {Ashoori},\ and\ \citenamefont {Jarillo-Herrero}}]{cao2018correlated}%
  \BibitemOpen
  \bibfield  {author} {\bibinfo {author} {\bibfnamefont {Y.}~\bibnamefont {Cao}}, \bibinfo {author} {\bibfnamefont {V.}~\bibnamefont {Fatemi}}, \bibinfo {author} {\bibfnamefont {A.}~\bibnamefont {Demir}}, \bibinfo {author} {\bibfnamefont {S.}~\bibnamefont {Fang}}, \bibinfo {author} {\bibfnamefont {S.~L.}\ \bibnamefont {Tomarken}}, \bibinfo {author} {\bibfnamefont {J.~Y.}\ \bibnamefont {Luo}}, \bibinfo {author} {\bibfnamefont {J.~D.}\ \bibnamefont {Sanchez-Yamagishi}}, \bibinfo {author} {\bibfnamefont {K.}~\bibnamefont {Watanabe}}, \bibinfo {author} {\bibfnamefont {T.}~\bibnamefont {Taniguchi}}, \bibinfo {author} {\bibfnamefont {E.}~\bibnamefont {Kaxiras}}, \bibinfo {author} {\bibfnamefont {R.~C.}\ \bibnamefont {Ashoori}},\ and\ \bibinfo {author} {\bibfnamefont {P.}~\bibnamefont {Jarillo-Herrero}},\ }\bibfield  {title} {\bibinfo {title} {Correlated insulator behaviour at half-filling in magic-angle graphene superlattices},\ }\href {https://doi.org/10.1038/nature26154} {\bibfield  {journal} {\bibinfo  {journal}
  {Nature}\ }\textbf {\bibinfo {volume} {556}},\ \bibinfo {pages} {80} (\bibinfo {year} {2018}{\natexlab{a}})}\BibitemShut {NoStop}%
\bibitem [{\citenamefont {Tarnopolsky}\ \emph {et~al.}(2019)\citenamefont {Tarnopolsky}, \citenamefont {Kruchkov},\ and\ \citenamefont {Vishwanath}}]{tarnopolsky2019origin}%
  \BibitemOpen
  \bibfield  {author} {\bibinfo {author} {\bibfnamefont {G.}~\bibnamefont {Tarnopolsky}}, \bibinfo {author} {\bibfnamefont {A.~J.}\ \bibnamefont {Kruchkov}},\ and\ \bibinfo {author} {\bibfnamefont {A.}~\bibnamefont {Vishwanath}},\ }\bibfield  {title} {\bibinfo {title} {Origin of magic angles in twisted bilayer graphene},\ }\href {https://doi.org/10.1103/PhysRevLett.122.106405} {\bibfield  {journal} {\bibinfo  {journal} {Physical Review Letters}\ }\textbf {\bibinfo {volume} {122}},\ \bibinfo {pages} {106405} (\bibinfo {year} {2019})}\BibitemShut {NoStop}%
\bibitem [{\citenamefont {Cao}\ \emph {et~al.}(2018{\natexlab{b}})\citenamefont {Cao}, \citenamefont {Fatemi}, \citenamefont {Fang}, \citenamefont {Watanabe}, \citenamefont {Taniguchi}, \citenamefont {Kaxiras},\ and\ \citenamefont {Jarillo-Herrero}}]{cao2018unconventional}%
  \BibitemOpen
  \bibfield  {author} {\bibinfo {author} {\bibfnamefont {Y.}~\bibnamefont {Cao}}, \bibinfo {author} {\bibfnamefont {V.}~\bibnamefont {Fatemi}}, \bibinfo {author} {\bibfnamefont {S.}~\bibnamefont {Fang}}, \bibinfo {author} {\bibfnamefont {K.}~\bibnamefont {Watanabe}}, \bibinfo {author} {\bibfnamefont {T.}~\bibnamefont {Taniguchi}}, \bibinfo {author} {\bibfnamefont {E.}~\bibnamefont {Kaxiras}},\ and\ \bibinfo {author} {\bibfnamefont {P.}~\bibnamefont {Jarillo-Herrero}},\ }\bibfield  {title} {\bibinfo {title} {Unconventional superconductivity in magic-angle graphene superlattices},\ }\href {https://doi.org/10.1038/nature26160} {\bibfield  {journal} {\bibinfo  {journal} {Nature}\ }\textbf {\bibinfo {volume} {556}},\ \bibinfo {pages} {43} (\bibinfo {year} {2018}{\natexlab{b}})}\BibitemShut {NoStop}%
\bibitem [{\citenamefont {Wu}\ \emph {et~al.}(2018)\citenamefont {Wu}, \citenamefont {MacDonald},\ and\ \citenamefont {Martin}}]{wu2018theory}%
  \BibitemOpen
  \bibfield  {author} {\bibinfo {author} {\bibfnamefont {F.}~\bibnamefont {Wu}}, \bibinfo {author} {\bibfnamefont {A.~H.}\ \bibnamefont {MacDonald}},\ and\ \bibinfo {author} {\bibfnamefont {I.}~\bibnamefont {Martin}},\ }\bibfield  {title} {\bibinfo {title} {Theory of phonon-mediated superconductivity in twisted bilayer graphene},\ }\href {https://doi.org/10.1103/PhysRevLett.121.257001} {\bibfield  {journal} {\bibinfo  {journal} {Physical Review Letters}\ }\textbf {\bibinfo {volume} {121}},\ \bibinfo {pages} {257001} (\bibinfo {year} {2018})}\BibitemShut {NoStop}%
\bibitem [{\citenamefont {Isobe}\ \emph {et~al.}(2018)\citenamefont {Isobe}, \citenamefont {Yuan},\ and\ \citenamefont {Fu}}]{isobe2018unconventional}%
  \BibitemOpen
  \bibfield  {author} {\bibinfo {author} {\bibfnamefont {H.}~\bibnamefont {Isobe}}, \bibinfo {author} {\bibfnamefont {N.~F.~Q.}\ \bibnamefont {Yuan}},\ and\ \bibinfo {author} {\bibfnamefont {L.}~\bibnamefont {Fu}},\ }\bibfield  {title} {\bibinfo {title} {Unconventional superconductivity and density waves in twisted bilayer graphene},\ }\href {https://doi.org/10.1103/PhysRevX.8.041041} {\bibfield  {journal} {\bibinfo  {journal} {Physical Review X}\ }\textbf {\bibinfo {volume} {8}},\ \bibinfo {pages} {041041} (\bibinfo {year} {2018})}\BibitemShut {NoStop}%
\bibitem [{\citenamefont {Lian}\ \emph {et~al.}(2019)\citenamefont {Lian}, \citenamefont {Wang},\ and\ \citenamefont {Bernevig}}]{lian2019twisted}%
  \BibitemOpen
  \bibfield  {author} {\bibinfo {author} {\bibfnamefont {B.}~\bibnamefont {Lian}}, \bibinfo {author} {\bibfnamefont {Z.}~\bibnamefont {Wang}},\ and\ \bibinfo {author} {\bibfnamefont {B.~A.}\ \bibnamefont {Bernevig}},\ }\bibfield  {title} {\bibinfo {title} {Twisted bilayer graphene: A phonon-driven superconductor},\ }\href {https://doi.org/10.1103/PhysRevLett.122.257002} {\bibfield  {journal} {\bibinfo  {journal} {Physical Review Letters}\ }\textbf {\bibinfo {volume} {122}},\ \bibinfo {pages} {257002} (\bibinfo {year} {2019})}\BibitemShut {NoStop}%
\bibitem [{\citenamefont {Cai}\ \emph {et~al.}(2023)\citenamefont {Cai}, \citenamefont {Anderson}, \citenamefont {Wang}, \citenamefont {Zhang}, \citenamefont {Liu}, \citenamefont {Holtzmann}, \citenamefont {Zhang}, \citenamefont {Fan}, \citenamefont {Taniguchi}, \citenamefont {Watanabe}, \citenamefont {Ran}, \citenamefont {Cao}, \citenamefont {Fu}, \citenamefont {Xiao}, \citenamefont {Yao},\ and\ \citenamefont {Xu}}]{cai23nature}%
  \BibitemOpen
  \bibfield  {author} {\bibinfo {author} {\bibfnamefont {J.}~\bibnamefont {Cai}}, \bibinfo {author} {\bibfnamefont {E.}~\bibnamefont {Anderson}}, \bibinfo {author} {\bibfnamefont {C.}~\bibnamefont {Wang}}, \bibinfo {author} {\bibfnamefont {X.}~\bibnamefont {Zhang}}, \bibinfo {author} {\bibfnamefont {X.}~\bibnamefont {Liu}}, \bibinfo {author} {\bibfnamefont {W.}~\bibnamefont {Holtzmann}}, \bibinfo {author} {\bibfnamefont {Y.}~\bibnamefont {Zhang}}, \bibinfo {author} {\bibfnamefont {F.}~\bibnamefont {Fan}}, \bibinfo {author} {\bibfnamefont {T.}~\bibnamefont {Taniguchi}}, \bibinfo {author} {\bibfnamefont {K.}~\bibnamefont {Watanabe}}, \bibinfo {author} {\bibfnamefont {Y.}~\bibnamefont {Ran}}, \bibinfo {author} {\bibfnamefont {T.}~\bibnamefont {Cao}}, \bibinfo {author} {\bibfnamefont {L.}~\bibnamefont {Fu}}, \bibinfo {author} {\bibfnamefont {D.}~\bibnamefont {Xiao}}, \bibinfo {author} {\bibfnamefont {W.}~\bibnamefont {Yao}},\ and\ \bibinfo {author} {\bibfnamefont {X.}~\bibnamefont {Xu}},\ }\bibfield  {title}
  {\bibinfo {title} {Signatures of fractional quantum anomalous hall states in twisted ${\mathrm{mote}}_{2}$},\ }\href {https://doi.org/10.1038/s41586-023-06289-w} {\bibfield  {journal} {\bibinfo  {journal} {Nature}\ }\textbf {\bibinfo {volume} {622}},\ \bibinfo {pages} {63} (\bibinfo {year} {2023})}\BibitemShut {NoStop}%
\bibitem [{\citenamefont {Park}\ \emph {et~al.}(2023)\citenamefont {Park}, \citenamefont {Cai}, \citenamefont {Anderson}, \citenamefont {Zhang}, \citenamefont {Zhu}, \citenamefont {Liu}, \citenamefont {Wang}, \citenamefont {Holtzmann}, \citenamefont {Hu}, \citenamefont {Liu}, \citenamefont {Taniguchi}, \citenamefont {Watanabe}, \citenamefont {Chu}, \citenamefont {Cao}, \citenamefont {Fu}, \citenamefont {Yao}, \citenamefont {Chang}, \citenamefont {Cobden}, \citenamefont {Xiao},\ and\ \citenamefont {Xu}}]{park23nature}%
  \BibitemOpen
  \bibfield  {author} {\bibinfo {author} {\bibfnamefont {H.}~\bibnamefont {Park}}, \bibinfo {author} {\bibfnamefont {J.}~\bibnamefont {Cai}}, \bibinfo {author} {\bibfnamefont {E.}~\bibnamefont {Anderson}}, \bibinfo {author} {\bibfnamefont {Y.}~\bibnamefont {Zhang}}, \bibinfo {author} {\bibfnamefont {J.}~\bibnamefont {Zhu}}, \bibinfo {author} {\bibfnamefont {X.}~\bibnamefont {Liu}}, \bibinfo {author} {\bibfnamefont {C.}~\bibnamefont {Wang}}, \bibinfo {author} {\bibfnamefont {W.}~\bibnamefont {Holtzmann}}, \bibinfo {author} {\bibfnamefont {C.}~\bibnamefont {Hu}}, \bibinfo {author} {\bibfnamefont {Z.}~\bibnamefont {Liu}}, \bibinfo {author} {\bibfnamefont {T.}~\bibnamefont {Taniguchi}}, \bibinfo {author} {\bibfnamefont {K.}~\bibnamefont {Watanabe}}, \bibinfo {author} {\bibfnamefont {J.-H.}\ \bibnamefont {Chu}}, \bibinfo {author} {\bibfnamefont {T.}~\bibnamefont {Cao}}, \bibinfo {author} {\bibfnamefont {L.}~\bibnamefont {Fu}}, \bibinfo {author} {\bibfnamefont {W.}~\bibnamefont {Yao}}, \bibinfo {author}
  {\bibfnamefont {C.-Z.}\ \bibnamefont {Chang}}, \bibinfo {author} {\bibfnamefont {D.}~\bibnamefont {Cobden}}, \bibinfo {author} {\bibfnamefont {D.}~\bibnamefont {Xiao}},\ and\ \bibinfo {author} {\bibfnamefont {X.}~\bibnamefont {Xu}},\ }\bibfield  {title} {\bibinfo {title} {Observation of fractionally quantized anomalous hall effect},\ }\href {https://doi.org/10.1038/s41586-023-06536-0} {\bibfield  {journal} {\bibinfo  {journal} {Nature}\ }\textbf {\bibinfo {volume} {622}},\ \bibinfo {pages} {74} (\bibinfo {year} {2023})}\BibitemShut {NoStop}%
\bibitem [{\citenamefont {Xu}\ \emph {et~al.}(2023)\citenamefont {Xu}, \citenamefont {Sun}, \citenamefont {Jia}, \citenamefont {Liu}, \citenamefont {Xu}, \citenamefont {Li}, \citenamefont {Gu}, \citenamefont {Watanabe}, \citenamefont {Taniguchi}, \citenamefont {Tong}, \citenamefont {Jia}, \citenamefont {Shi}, \citenamefont {Jiang}, \citenamefont {Zhang}, \citenamefont {Liu},\ and\ \citenamefont {Li}}]{xu2023observation}%
  \BibitemOpen
  \bibfield  {author} {\bibinfo {author} {\bibfnamefont {F.}~\bibnamefont {Xu}}, \bibinfo {author} {\bibfnamefont {Z.}~\bibnamefont {Sun}}, \bibinfo {author} {\bibfnamefont {T.}~\bibnamefont {Jia}}, \bibinfo {author} {\bibfnamefont {C.}~\bibnamefont {Liu}}, \bibinfo {author} {\bibfnamefont {C.}~\bibnamefont {Xu}}, \bibinfo {author} {\bibfnamefont {C.}~\bibnamefont {Li}}, \bibinfo {author} {\bibfnamefont {Y.}~\bibnamefont {Gu}}, \bibinfo {author} {\bibfnamefont {K.}~\bibnamefont {Watanabe}}, \bibinfo {author} {\bibfnamefont {T.}~\bibnamefont {Taniguchi}}, \bibinfo {author} {\bibfnamefont {B.}~\bibnamefont {Tong}}, \bibinfo {author} {\bibfnamefont {J.}~\bibnamefont {Jia}}, \bibinfo {author} {\bibfnamefont {Z.}~\bibnamefont {Shi}}, \bibinfo {author} {\bibfnamefont {S.}~\bibnamefont {Jiang}}, \bibinfo {author} {\bibfnamefont {Y.}~\bibnamefont {Zhang}}, \bibinfo {author} {\bibfnamefont {X.}~\bibnamefont {Liu}},\ and\ \bibinfo {author} {\bibfnamefont {T.}~\bibnamefont {Li}},\ }\bibfield  {title} {\bibinfo {title}
  {Observation of integer and fractional quantum anomalous hall effects in twisted bilayer ${\mathrm{mote}}_{2}$},\ }\href {https://doi.org/10.1103/PhysRevX.13.031037} {\bibfield  {journal} {\bibinfo  {journal} {Physical Review X}\ }\textbf {\bibinfo {volume} {13}},\ \bibinfo {pages} {031037} (\bibinfo {year} {2023})}\BibitemShut {NoStop}%
\bibitem [{\citenamefont {Lu}\ \emph {et~al.}(2024)\citenamefont {Lu}, \citenamefont {Han}, \citenamefont {Yao}, \citenamefont {Reddy}, \citenamefont {Yang}, \citenamefont {Seo}, \citenamefont {Watanabe}, \citenamefont {Taniguchi}, \citenamefont {Fu},\ and\ \citenamefont {Ju}}]{lu_2024}%
  \BibitemOpen
  \bibfield  {author} {\bibinfo {author} {\bibfnamefont {Z.}~\bibnamefont {Lu}}, \bibinfo {author} {\bibfnamefont {T.}~\bibnamefont {Han}}, \bibinfo {author} {\bibfnamefont {Y.}~\bibnamefont {Yao}}, \bibinfo {author} {\bibfnamefont {A.~P.}\ \bibnamefont {Reddy}}, \bibinfo {author} {\bibfnamefont {J.}~\bibnamefont {Yang}}, \bibinfo {author} {\bibfnamefont {J.}~\bibnamefont {Seo}}, \bibinfo {author} {\bibfnamefont {K.}~\bibnamefont {Watanabe}}, \bibinfo {author} {\bibfnamefont {T.}~\bibnamefont {Taniguchi}}, \bibinfo {author} {\bibfnamefont {L.}~\bibnamefont {Fu}},\ and\ \bibinfo {author} {\bibfnamefont {L.}~\bibnamefont {Ju}},\ }\bibfield  {title} {\bibinfo {title} {Fractional quantum anomalous hall effect in multilayer graphene},\ }\href {https://doi.org/10.1038/s41586-023-07010-7} {\bibfield  {journal} {\bibinfo  {journal} {Nature}\ }\textbf {\bibinfo {volume} {626}},\ \bibinfo {pages} {759} (\bibinfo {year} {2024})}\BibitemShut {NoStop}%
\bibitem [{\citenamefont {Sunku}\ \emph {et~al.}(2018)\citenamefont {Sunku}, \citenamefont {Ni}, \citenamefont {Jiang}, \citenamefont {Yoo}, \citenamefont {Sternbach}, \citenamefont {McLeod}, \citenamefont {Stauber}, \citenamefont {Xiong}, \citenamefont {Taniguchi}, \citenamefont {Watanabe}, \citenamefont {Kim}, \citenamefont {Fogler},\ and\ \citenamefont {Basov}}]{18science}%
  \BibitemOpen
  \bibfield  {author} {\bibinfo {author} {\bibfnamefont {S.}~\bibnamefont {Sunku}}, \bibinfo {author} {\bibfnamefont {G.}~\bibnamefont {Ni}}, \bibinfo {author} {\bibfnamefont {B.-Y.}\ \bibnamefont {Jiang}}, \bibinfo {author} {\bibfnamefont {H.}~\bibnamefont {Yoo}}, \bibinfo {author} {\bibfnamefont {A.}~\bibnamefont {Sternbach}}, \bibinfo {author} {\bibfnamefont {A.}~\bibnamefont {McLeod}}, \bibinfo {author} {\bibfnamefont {T.}~\bibnamefont {Stauber}}, \bibinfo {author} {\bibfnamefont {L.}~\bibnamefont {Xiong}}, \bibinfo {author} {\bibfnamefont {T.}~\bibnamefont {Taniguchi}}, \bibinfo {author} {\bibfnamefont {K.}~\bibnamefont {Watanabe}}, \bibinfo {author} {\bibfnamefont {P.}~\bibnamefont {Kim}}, \bibinfo {author} {\bibfnamefont {M.}~\bibnamefont {Fogler}},\ and\ \bibinfo {author} {\bibfnamefont {D.}~\bibnamefont {Basov}},\ }\bibfield  {title} {\bibinfo {title} {Photonic crystals for nano-light in moir{\'e} graphene superlattices},\ }\href {https://doi.org/10.1126/science.aau514} {\bibfield  {journal}
  {\bibinfo  {journal} {Science}\ }\textbf {\bibinfo {volume} {362}},\ \bibinfo {pages} {1153} (\bibinfo {year} {2018})}\BibitemShut {NoStop}%
\bibitem [{\citenamefont {Wang}\ \emph {et~al.}(2020)\citenamefont {Wang}, \citenamefont {Zheng}, \citenamefont {Chen}, \citenamefont {Huang}, \citenamefont {Kartashov}, \citenamefont {Torner}, \citenamefont {Konotop},\ and\ \citenamefont {Ye}}]{wang2020localization}%
  \BibitemOpen
  \bibfield  {author} {\bibinfo {author} {\bibfnamefont {P.}~\bibnamefont {Wang}}, \bibinfo {author} {\bibfnamefont {Y.}~\bibnamefont {Zheng}}, \bibinfo {author} {\bibfnamefont {X.}~\bibnamefont {Chen}}, \bibinfo {author} {\bibfnamefont {C.}~\bibnamefont {Huang}}, \bibinfo {author} {\bibfnamefont {Y.~V.}\ \bibnamefont {Kartashov}}, \bibinfo {author} {\bibfnamefont {L.}~\bibnamefont {Torner}}, \bibinfo {author} {\bibfnamefont {V.~V.}\ \bibnamefont {Konotop}},\ and\ \bibinfo {author} {\bibfnamefont {F.}~\bibnamefont {Ye}},\ }\bibfield  {title} {\bibinfo {title} {Localization and delocalization of light in photonic moir{\'e} lattices},\ }\href {https://doi.org/10.1038/s41586-019-1851-6} {\bibfield  {journal} {\bibinfo  {journal} {Nature}\ }\textbf {\bibinfo {volume} {577}},\ \bibinfo {pages} {42} (\bibinfo {year} {2020})}\BibitemShut {NoStop}%
\bibitem [{\citenamefont {Mao}\ \emph {et~al.}(2021)\citenamefont {Mao}, \citenamefont {Shao}, \citenamefont {Luan}, \citenamefont {Wang},\ and\ \citenamefont {Ma}}]{mao2021magic}%
  \BibitemOpen
  \bibfield  {author} {\bibinfo {author} {\bibfnamefont {X.-R.}\ \bibnamefont {Mao}}, \bibinfo {author} {\bibfnamefont {Z.-K.}\ \bibnamefont {Shao}}, \bibinfo {author} {\bibfnamefont {H.-Y.}\ \bibnamefont {Luan}}, \bibinfo {author} {\bibfnamefont {S.-L.}\ \bibnamefont {Wang}},\ and\ \bibinfo {author} {\bibfnamefont {R.-M.}\ \bibnamefont {Ma}},\ }\bibfield  {title} {\bibinfo {title} {Magic-angle lasers in nanostructured moir{\'e} superlattice},\ }\href {https://doi.org/10.1038/s41565-021-00956-7} {\bibfield  {journal} {\bibinfo  {journal} {Nature nanotechnology}\ }\textbf {\bibinfo {volume} {16}},\ \bibinfo {pages} {1099} (\bibinfo {year} {2021})}\BibitemShut {NoStop}%
\bibitem [{\citenamefont {Dong}\ \emph {et~al.}(2021)\citenamefont {Dong}, \citenamefont {Zhang}, \citenamefont {Li}, \citenamefont {Wang}, \citenamefont {Yang}, \citenamefont {Rho}, \citenamefont {Wang}, \citenamefont {Grigoropoulos}, \citenamefont {Wu},\ and\ \citenamefont {Yao}}]{Dong21PRL}%
  \BibitemOpen
  \bibfield  {author} {\bibinfo {author} {\bibfnamefont {K.}~\bibnamefont {Dong}}, \bibinfo {author} {\bibfnamefont {T.}~\bibnamefont {Zhang}}, \bibinfo {author} {\bibfnamefont {J.}~\bibnamefont {Li}}, \bibinfo {author} {\bibfnamefont {Q.}~\bibnamefont {Wang}}, \bibinfo {author} {\bibfnamefont {F.}~\bibnamefont {Yang}}, \bibinfo {author} {\bibfnamefont {Y.}~\bibnamefont {Rho}}, \bibinfo {author} {\bibfnamefont {D.}~\bibnamefont {Wang}}, \bibinfo {author} {\bibfnamefont {C.~P.}\ \bibnamefont {Grigoropoulos}}, \bibinfo {author} {\bibfnamefont {J.}~\bibnamefont {Wu}},\ and\ \bibinfo {author} {\bibfnamefont {J.}~\bibnamefont {Yao}},\ }\bibfield  {title} {\bibinfo {title} {Flat bands in magic-angle bilayer photonic crystals at small twists},\ }\href {https://doi.org/10.1103/PhysRevLett.126.223601} {\bibfield  {journal} {\bibinfo  {journal} {Physical Review Letters}\ }\textbf {\bibinfo {volume} {126}},\ \bibinfo {pages} {223601} (\bibinfo {year} {2021})}\BibitemShut {NoStop}%
\bibitem [{\citenamefont {Lou}\ \emph {et~al.}(2022)\citenamefont {Lou}, \citenamefont {Wang}, \citenamefont {Rodríguez}, \citenamefont {Cappelli},\ and\ \citenamefont {Fan}}]{Lou_2022}%
  \BibitemOpen
  \bibfield  {author} {\bibinfo {author} {\bibfnamefont {B.}~\bibnamefont {Lou}}, \bibinfo {author} {\bibfnamefont {B.}~\bibnamefont {Wang}}, \bibinfo {author} {\bibfnamefont {J.~A.}\ \bibnamefont {Rodríguez}}, \bibinfo {author} {\bibfnamefont {M.}~\bibnamefont {Cappelli}},\ and\ \bibinfo {author} {\bibfnamefont {S.}~\bibnamefont {Fan}},\ }\bibfield  {title} {\bibinfo {title} {Tunable guided resonance in twisted bilayer photonic crystal},\ }\href {https://doi.org/10.1126/sciadv.add4339} {\bibfield  {journal} {\bibinfo  {journal} {Science Advances}\ }\textbf {\bibinfo {volume} {8}},\ \bibinfo {pages} {eadd4339} (\bibinfo {year} {2022})}\BibitemShut {NoStop}%
\bibitem [{\citenamefont {Tang}\ \emph {et~al.}(2023)\citenamefont {Tang}, \citenamefont {Lou}, \citenamefont {Du}, \citenamefont {Zhang}, \citenamefont {Ni}, \citenamefont {Xu}, \citenamefont {Jin}, \citenamefont {Fan},\ and\ \citenamefont {Mazur}}]{Tang_2023}%
  \BibitemOpen
  \bibfield  {author} {\bibinfo {author} {\bibfnamefont {H.}~\bibnamefont {Tang}}, \bibinfo {author} {\bibfnamefont {B.}~\bibnamefont {Lou}}, \bibinfo {author} {\bibfnamefont {F.}~\bibnamefont {Du}}, \bibinfo {author} {\bibfnamefont {M.}~\bibnamefont {Zhang}}, \bibinfo {author} {\bibfnamefont {X.}~\bibnamefont {Ni}}, \bibinfo {author} {\bibfnamefont {W.}~\bibnamefont {Xu}}, \bibinfo {author} {\bibfnamefont {R.}~\bibnamefont {Jin}}, \bibinfo {author} {\bibfnamefont {S.}~\bibnamefont {Fan}},\ and\ \bibinfo {author} {\bibfnamefont {E.}~\bibnamefont {Mazur}},\ }\bibfield  {title} {\bibinfo {title} {Experimental probe of twist angle–dependent band structure of on-chip optical bilayer photonic crystal},\ }\href {https://doi.org/10.1126/sciadv.adh8498} {\bibfield  {journal} {\bibinfo  {journal} {Science Advances}\ }\textbf {\bibinfo {volume} {9}},\ \bibinfo {pages} {eadh8498} (\bibinfo {year} {2023})}\BibitemShut {NoStop}%
\bibitem [{\citenamefont {Zheng}\ \emph {et~al.}(2020)\citenamefont {Zheng}, \citenamefont {Sun}, \citenamefont {Huang}, \citenamefont {Jiang}, \citenamefont {Zhan}, \citenamefont {Ke}, \citenamefont {Chen},\ and\ \citenamefont {Deng}}]{zheng2020phonon}%
  \BibitemOpen
  \bibfield  {author} {\bibinfo {author} {\bibfnamefont {Z.}~\bibnamefont {Zheng}}, \bibinfo {author} {\bibfnamefont {F.}~\bibnamefont {Sun}}, \bibinfo {author} {\bibfnamefont {W.}~\bibnamefont {Huang}}, \bibinfo {author} {\bibfnamefont {J.}~\bibnamefont {Jiang}}, \bibinfo {author} {\bibfnamefont {R.}~\bibnamefont {Zhan}}, \bibinfo {author} {\bibfnamefont {Y.}~\bibnamefont {Ke}}, \bibinfo {author} {\bibfnamefont {H.}~\bibnamefont {Chen}},\ and\ \bibinfo {author} {\bibfnamefont {S.}~\bibnamefont {Deng}},\ }\bibfield  {title} {\bibinfo {title} {Phonon polaritons in twisted double-layers of hyperbolic van der waals crystals},\ }\href {https://doi.org/10.1021/acs.nanolett.0c01627} {\bibfield  {journal} {\bibinfo  {journal} {Nano letters}\ }\textbf {\bibinfo {volume} {20}},\ \bibinfo {pages} {5301} (\bibinfo {year} {2020})}\BibitemShut {NoStop}%
\bibitem [{\citenamefont {Duan}\ \emph {et~al.}(2020)\citenamefont {Duan}, \citenamefont {Capote-Robayna}, \citenamefont {Taboada-Guti{\'e}rrez}, \citenamefont {{\'A}lvarez-P{\'e}rez}, \citenamefont {Prieto}, \citenamefont {Mart{\'\i}n-S{\'a}nchez}, \citenamefont {Nikitin},\ and\ \citenamefont {Alonso-Gonz{\'a}lez}}]{duan2020twisted}%
  \BibitemOpen
  \bibfield  {author} {\bibinfo {author} {\bibfnamefont {J.}~\bibnamefont {Duan}}, \bibinfo {author} {\bibfnamefont {N.}~\bibnamefont {Capote-Robayna}}, \bibinfo {author} {\bibfnamefont {J.}~\bibnamefont {Taboada-Guti{\'e}rrez}}, \bibinfo {author} {\bibfnamefont {G.}~\bibnamefont {{\'A}lvarez-P{\'e}rez}}, \bibinfo {author} {\bibfnamefont {I.}~\bibnamefont {Prieto}}, \bibinfo {author} {\bibfnamefont {J.}~\bibnamefont {Mart{\'\i}n-S{\'a}nchez}}, \bibinfo {author} {\bibfnamefont {A.~Y.}\ \bibnamefont {Nikitin}},\ and\ \bibinfo {author} {\bibfnamefont {P.}~\bibnamefont {Alonso-Gonz{\'a}lez}},\ }\bibfield  {title} {\bibinfo {title} {Twisted nano-optics: manipulating light at the nanoscale with twisted phonon polaritonic slabs},\ }\href {https://doi.org/10.1021/acs.nanolett.0c01673} {\bibfield  {journal} {\bibinfo  {journal} {Nano Letters}\ }\textbf {\bibinfo {volume} {20}},\ \bibinfo {pages} {5323} (\bibinfo {year} {2020})}\BibitemShut {NoStop}%
\bibitem [{\citenamefont {Hu}\ \emph {et~al.}(2020)\citenamefont {Hu}, \citenamefont {Ou}, \citenamefont {Si}, \citenamefont {Wu}, \citenamefont {Wu}, \citenamefont {Dai}, \citenamefont {Krasnok}, \citenamefont {Mazor}, \citenamefont {Zhang}, \citenamefont {Bao}, \citenamefont {Qiu},\ and\ \citenamefont {Alù}}]{hu2020topological}%
  \BibitemOpen
  \bibfield  {author} {\bibinfo {author} {\bibfnamefont {G.}~\bibnamefont {Hu}}, \bibinfo {author} {\bibfnamefont {Q.}~\bibnamefont {Ou}}, \bibinfo {author} {\bibfnamefont {G.}~\bibnamefont {Si}}, \bibinfo {author} {\bibfnamefont {Y.}~\bibnamefont {Wu}}, \bibinfo {author} {\bibfnamefont {J.}~\bibnamefont {Wu}}, \bibinfo {author} {\bibfnamefont {Z.}~\bibnamefont {Dai}}, \bibinfo {author} {\bibfnamefont {A.}~\bibnamefont {Krasnok}}, \bibinfo {author} {\bibfnamefont {Y.}~\bibnamefont {Mazor}}, \bibinfo {author} {\bibfnamefont {Q.}~\bibnamefont {Zhang}}, \bibinfo {author} {\bibfnamefont {Q.}~\bibnamefont {Bao}}, \bibinfo {author} {\bibfnamefont {C.-W.}\ \bibnamefont {Qiu}},\ and\ \bibinfo {author} {\bibfnamefont {A.}~\bibnamefont {Alù}},\ }\bibfield  {title} {\bibinfo {title} {Topological polaritons and photonic magic angles in twisted $\alpha$-moo3 bilayers},\ }\href {https://doi.org/10.1038/s41586-020-2359-9} {\bibfield  {journal} {\bibinfo  {journal} {Nature}\ }\textbf {\bibinfo {volume} {582}},\ \bibinfo
  {pages} {209} (\bibinfo {year} {2020})}\BibitemShut {NoStop}%
\bibitem [{\citenamefont {Chen}\ \emph {et~al.}(2020)\citenamefont {Chen}, \citenamefont {Lin}, \citenamefont {Dinh}, \citenamefont {Zheng}, \citenamefont {Shen}, \citenamefont {Ma}, \citenamefont {Chen}, \citenamefont {Jarillo-Herrero},\ and\ \citenamefont {Dai}}]{chen2020configurable}%
  \BibitemOpen
  \bibfield  {author} {\bibinfo {author} {\bibfnamefont {M.}~\bibnamefont {Chen}}, \bibinfo {author} {\bibfnamefont {X.}~\bibnamefont {Lin}}, \bibinfo {author} {\bibfnamefont {T.~H.}\ \bibnamefont {Dinh}}, \bibinfo {author} {\bibfnamefont {Z.}~\bibnamefont {Zheng}}, \bibinfo {author} {\bibfnamefont {J.}~\bibnamefont {Shen}}, \bibinfo {author} {\bibfnamefont {Q.}~\bibnamefont {Ma}}, \bibinfo {author} {\bibfnamefont {H.}~\bibnamefont {Chen}}, \bibinfo {author} {\bibfnamefont {P.}~\bibnamefont {Jarillo-Herrero}},\ and\ \bibinfo {author} {\bibfnamefont {S.}~\bibnamefont {Dai}},\ }\bibfield  {title} {\bibinfo {title} {Configurable phonon polaritons in twisted $\alpha$-moo3},\ }\href {https://doi.org/10.1038/s41563-020-0732-6} {\bibfield  {journal} {\bibinfo  {journal} {Nature materials}\ }\textbf {\bibinfo {volume} {19}},\ \bibinfo {pages} {1307} (\bibinfo {year} {2020})}\BibitemShut {NoStop}%
\bibitem [{\citenamefont {Yao}\ \emph {et~al.}(2024)\citenamefont {Yao}, \citenamefont {Ye}, \citenamefont {Fu}, \citenamefont {Wang}, \citenamefont {He}, \citenamefont {Lu}, \citenamefont {Deng}, \citenamefont {Huang}, \citenamefont {Ke},\ and\ \citenamefont {Liu}}]{Yao24PRL}%
  \BibitemOpen
  \bibfield  {author} {\bibinfo {author} {\bibfnamefont {D.}~\bibnamefont {Yao}}, \bibinfo {author} {\bibfnamefont {L.}~\bibnamefont {Ye}}, \bibinfo {author} {\bibfnamefont {Z.}~\bibnamefont {Fu}}, \bibinfo {author} {\bibfnamefont {Q.}~\bibnamefont {Wang}}, \bibinfo {author} {\bibfnamefont {H.}~\bibnamefont {He}}, \bibinfo {author} {\bibfnamefont {J.}~\bibnamefont {Lu}}, \bibinfo {author} {\bibfnamefont {W.}~\bibnamefont {Deng}}, \bibinfo {author} {\bibfnamefont {X.}~\bibnamefont {Huang}}, \bibinfo {author} {\bibfnamefont {M.}~\bibnamefont {Ke}},\ and\ \bibinfo {author} {\bibfnamefont {Z.}~\bibnamefont {Liu}},\ }\bibfield  {title} {\bibinfo {title} {Topological network modes in a twisted moir\'e phononic crystal},\ }\href {https://doi.org/10.1103/PhysRevLett.132.266602} {\bibfield  {journal} {\bibinfo  {journal} {Physical Review Letters}\ }\textbf {\bibinfo {volume} {132}},\ \bibinfo {pages} {266602} (\bibinfo {year} {2024})}\BibitemShut {NoStop}%
\bibitem [{\citenamefont {Ma}\ \emph {et~al.}(2023)\citenamefont {Ma}, \citenamefont {Luan}, \citenamefont {Zhao}, \citenamefont {Mao}, \citenamefont {Wang}, \citenamefont {Ouyang},\ and\ \citenamefont {Shao}}]{23FR}%
  \BibitemOpen
  \bibfield  {author} {\bibinfo {author} {\bibfnamefont {R.-M.}\ \bibnamefont {Ma}}, \bibinfo {author} {\bibfnamefont {H.-Y.}\ \bibnamefont {Luan}}, \bibinfo {author} {\bibfnamefont {Z.-W.}\ \bibnamefont {Zhao}}, \bibinfo {author} {\bibfnamefont {W.-Z.}\ \bibnamefont {Mao}}, \bibinfo {author} {\bibfnamefont {S.-L.}\ \bibnamefont {Wang}}, \bibinfo {author} {\bibfnamefont {Y.-H.}\ \bibnamefont {Ouyang}},\ and\ \bibinfo {author} {\bibfnamefont {Z.-K.}\ \bibnamefont {Shao}},\ }\bibfield  {title} {\bibinfo {title} {Twisted lattice nanocavity with theoretical quality factor exceeding 200 billion},\ }\href {https://doi.org/https://doi.org/10.1016/j.fmre.2022.11.004} {\bibfield  {journal} {\bibinfo  {journal} {Fundamental Research}\ }\textbf {\bibinfo {volume} {3}},\ \bibinfo {pages} {537} (\bibinfo {year} {2023})}\BibitemShut {NoStop}%
\bibitem [{\citenamefont {Luan}\ \emph {et~al.}(2023)\citenamefont {Luan}, \citenamefont {Ouyang}, \citenamefont {Zhao}, \citenamefont {Mao},\ and\ \citenamefont {Ma}}]{luan2023reconfigurable}%
  \BibitemOpen
  \bibfield  {author} {\bibinfo {author} {\bibfnamefont {H.-Y.}\ \bibnamefont {Luan}}, \bibinfo {author} {\bibfnamefont {Y.-H.}\ \bibnamefont {Ouyang}}, \bibinfo {author} {\bibfnamefont {Z.-W.}\ \bibnamefont {Zhao}}, \bibinfo {author} {\bibfnamefont {W.-Z.}\ \bibnamefont {Mao}},\ and\ \bibinfo {author} {\bibfnamefont {R.-M.}\ \bibnamefont {Ma}},\ }\bibfield  {title} {\bibinfo {title} {Reconfigurable moir{\'e} nanolaser arrays with phase synchronization},\ }\href {https://doi.org/10.1038/s41586-023-06789-9} {\bibfield  {journal} {\bibinfo  {journal} {Nature}\ }\textbf {\bibinfo {volume} {624}},\ \bibinfo {pages} {282} (\bibinfo {year} {2023})}\BibitemShut {NoStop}%
\bibitem [{\citenamefont {Spencer}\ \emph {et~al.}(2025)\citenamefont {Spencer}, \citenamefont {Coste}, \citenamefont {Ni}, \citenamefont {Park}, \citenamefont {Schaeper}, \citenamefont {Kim}, \citenamefont {Taniguchi}, \citenamefont {Watanabe}, \citenamefont {Toth}, \citenamefont {Zalogina}, \citenamefont {Tang},\ and\ \citenamefont {Aharonovichs}}]{spencer2025van}%
  \BibitemOpen
  \bibfield  {author} {\bibinfo {author} {\bibfnamefont {L.}~\bibnamefont {Spencer}}, \bibinfo {author} {\bibfnamefont {N.}~\bibnamefont {Coste}}, \bibinfo {author} {\bibfnamefont {X.}~\bibnamefont {Ni}}, \bibinfo {author} {\bibfnamefont {S.}~\bibnamefont {Park}}, \bibinfo {author} {\bibfnamefont {O.~C.}\ \bibnamefont {Schaeper}}, \bibinfo {author} {\bibfnamefont {Y.~D.}\ \bibnamefont {Kim}}, \bibinfo {author} {\bibfnamefont {T.}~\bibnamefont {Taniguchi}}, \bibinfo {author} {\bibfnamefont {K.}~\bibnamefont {Watanabe}}, \bibinfo {author} {\bibfnamefont {M.}~\bibnamefont {Toth}}, \bibinfo {author} {\bibfnamefont {A.}~\bibnamefont {Zalogina}}, \bibinfo {author} {\bibfnamefont {H.}~\bibnamefont {Tang}},\ and\ \bibinfo {author} {\bibfnamefont {I.}~\bibnamefont {Aharonovichs}},\ }\bibfield  {title} {\bibinfo {title} {A van der waals moir\'e bilayer photonic crystal cavity},\ }\href {https://arxiv.org/abs/2502.09839} {\bibfield  {journal} {\bibinfo  {journal} {e-print arXiv: 2502.09839}\ } (\bibinfo {year}
  {2025})}\BibitemShut {NoStop}%
\bibitem [{\citenamefont {Gonz{\'a}lez-Tudela}\ and\ \citenamefont {Cirac}(2019)}]{gonzalez2019cold}%
  \BibitemOpen
  \bibfield  {author} {\bibinfo {author} {\bibfnamefont {A.}~\bibnamefont {Gonz{\'a}lez-Tudela}}\ and\ \bibinfo {author} {\bibfnamefont {J.~I.}\ \bibnamefont {Cirac}},\ }\bibfield  {title} {\bibinfo {title} {Cold atoms in twisted-bilayer optical potentials},\ }\href {https://doi.org/10.1103/PhysRevA.100.053604} {\bibfield  {journal} {\bibinfo  {journal} {Physical Review A}\ }\textbf {\bibinfo {volume} {100}},\ \bibinfo {pages} {053604} (\bibinfo {year} {2019})}\BibitemShut {NoStop}%
\bibitem [{\citenamefont {Salamon}\ \emph {et~al.}(2020)\citenamefont {Salamon}, \citenamefont {Celi}, \citenamefont {Chhajlany}, \citenamefont {Fr\'erot}, \citenamefont {Lewenstein}, \citenamefont {Tarruell},\ and\ \citenamefont {Rakshit}}]{20PRL}%
  \BibitemOpen
  \bibfield  {author} {\bibinfo {author} {\bibfnamefont {T.}~\bibnamefont {Salamon}}, \bibinfo {author} {\bibfnamefont {A.}~\bibnamefont {Celi}}, \bibinfo {author} {\bibfnamefont {R.~W.}\ \bibnamefont {Chhajlany}}, \bibinfo {author} {\bibfnamefont {I.}~\bibnamefont {Fr\'erot}}, \bibinfo {author} {\bibfnamefont {M.}~\bibnamefont {Lewenstein}}, \bibinfo {author} {\bibfnamefont {L.}~\bibnamefont {Tarruell}},\ and\ \bibinfo {author} {\bibfnamefont {D.}~\bibnamefont {Rakshit}},\ }\bibfield  {title} {\bibinfo {title} {Simulating twistronics without a twist},\ }\href {https://doi.org/10.1103/PhysRevLett.125.030504} {\bibfield  {journal} {\bibinfo  {journal} {Physical Review Letters}\ }\textbf {\bibinfo {volume} {125}},\ \bibinfo {pages} {030504} (\bibinfo {year} {2020})}\BibitemShut {NoStop}%
\bibitem [{\citenamefont {Luo}\ and\ \citenamefont {Zhang}(2021)}]{Luo_2021}%
  \BibitemOpen
  \bibfield  {author} {\bibinfo {author} {\bibfnamefont {X.-W.}\ \bibnamefont {Luo}}\ and\ \bibinfo {author} {\bibfnamefont {C.}~\bibnamefont {Zhang}},\ }\bibfield  {title} {\bibinfo {title} {Spin-twisted optical lattices: Tunable flat bands and larkin-ovchinnikov superfluids},\ }\href {https://doi.org/10.1103/physrevlett.126.103201} {\bibfield  {journal} {\bibinfo  {journal} {Physical Review Letters}\ }\textbf {\bibinfo {volume} {126}},\ \bibinfo {pages} {103201} (\bibinfo {year} {2021})}\BibitemShut {NoStop}%
\bibitem [{\citenamefont {Paul}\ \emph {et~al.}(2023)\citenamefont {Paul}, \citenamefont {Recher},\ and\ \citenamefont {Santos}}]{Paul23PRA}%
  \BibitemOpen
  \bibfield  {author} {\bibinfo {author} {\bibfnamefont {G.~C.}\ \bibnamefont {Paul}}, \bibinfo {author} {\bibfnamefont {P.}~\bibnamefont {Recher}},\ and\ \bibinfo {author} {\bibfnamefont {L.}~\bibnamefont {Santos}},\ }\bibfield  {title} {\bibinfo {title} {Particle dynamics and ergodicity breaking in twisted-bilayer optical lattices},\ }\href {https://doi.org/10.1103/PhysRevA.108.053305} {\bibfield  {journal} {\bibinfo  {journal} {Physical Review A}\ }\textbf {\bibinfo {volume} {108}},\ \bibinfo {pages} {053305} (\bibinfo {year} {2023})}\BibitemShut {NoStop}%
\bibitem [{\citenamefont {Meng}\ \emph {et~al.}(2023)\citenamefont {Meng}, \citenamefont {Wang}, \citenamefont {Han}, \citenamefont {Liu}, \citenamefont {Wen}, \citenamefont {Gao}, \citenamefont {Wang}, \citenamefont {Chin},\ and\ \citenamefont {Zhang}}]{meng2023atomic}%
  \BibitemOpen
  \bibfield  {author} {\bibinfo {author} {\bibfnamefont {Z.}~\bibnamefont {Meng}}, \bibinfo {author} {\bibfnamefont {L.}~\bibnamefont {Wang}}, \bibinfo {author} {\bibfnamefont {W.}~\bibnamefont {Han}}, \bibinfo {author} {\bibfnamefont {F.}~\bibnamefont {Liu}}, \bibinfo {author} {\bibfnamefont {K.}~\bibnamefont {Wen}}, \bibinfo {author} {\bibfnamefont {C.}~\bibnamefont {Gao}}, \bibinfo {author} {\bibfnamefont {P.}~\bibnamefont {Wang}}, \bibinfo {author} {\bibfnamefont {C.}~\bibnamefont {Chin}},\ and\ \bibinfo {author} {\bibfnamefont {J.}~\bibnamefont {Zhang}},\ }\bibfield  {title} {\bibinfo {title} {Atomic bose--einstein condensate in twisted-bilayer optical lattices},\ }\href {https://doi.org/10.1038/s41586-023-05695-4} {\bibfield  {journal} {\bibinfo  {journal} {Nature}\ }\textbf {\bibinfo {volume} {615}},\ \bibinfo {pages} {231} (\bibinfo {year} {2023})}\BibitemShut {NoStop}%
\bibitem [{\citenamefont {Wan}\ \emph {et~al.}(2024)\citenamefont {Wan}, \citenamefont {Gao},\ and\ \citenamefont {Shi}}]{wan2024fractal}%
  \BibitemOpen
  \bibfield  {author} {\bibinfo {author} {\bibfnamefont {X.-T.}\ \bibnamefont {Wan}}, \bibinfo {author} {\bibfnamefont {C.}~\bibnamefont {Gao}},\ and\ \bibinfo {author} {\bibfnamefont {Z.-Y.}\ \bibnamefont {Shi}},\ }\bibfield  {title} {\bibinfo {title} {Fractal spectrum in twisted bilayer optical lattice},\ }\href {https://arxiv.org/abs/2404.08211} {\bibfield  {journal} {\bibinfo  {journal} {e-print arXiv: 2404.08211}\ } (\bibinfo {year} {2024})}\BibitemShut {NoStop}%
\bibitem [{\citenamefont {Zeng}\ \emph {et~al.}(2024)\citenamefont {Zeng}, \citenamefont {Zhu},\ and\ \citenamefont {He}}]{zeng2024dynamical}%
  \BibitemOpen
  \bibfield  {author} {\bibinfo {author} {\bibfnamefont {J.-H.}\ \bibnamefont {Zeng}}, \bibinfo {author} {\bibfnamefont {Q.}~\bibnamefont {Zhu}},\ and\ \bibinfo {author} {\bibfnamefont {L.}~\bibnamefont {He}},\ }\bibfield  {title} {\bibinfo {title} {Dynamical moir$\backslash$'e systems in twisted bilayer optical lattices},\ }\href {https://arxiv.org/abs/2405.20732} {\bibfield  {journal} {\bibinfo  {journal} {e-print arXiv: 2405.20732}\ } (\bibinfo {year} {2024})}\BibitemShut {NoStop}%
\bibitem [{\citenamefont {Tian}\ \emph {et~al.}(2024)\citenamefont {Tian}, \citenamefont {Zhang}, \citenamefont {Wu}, \citenamefont {Liu}, \citenamefont {Zhang}, \citenamefont {Li},\ and\ \citenamefont {Liu}}]{tian2024nonlinearity}%
  \BibitemOpen
  \bibfield  {author} {\bibinfo {author} {\bibfnamefont {R.}~\bibnamefont {Tian}}, \bibinfo {author} {\bibfnamefont {Y.}~\bibnamefont {Zhang}}, \bibinfo {author} {\bibfnamefont {T.}~\bibnamefont {Wu}}, \bibinfo {author} {\bibfnamefont {M.}~\bibnamefont {Liu}}, \bibinfo {author} {\bibfnamefont {Y.-C.}\ \bibnamefont {Zhang}}, \bibinfo {author} {\bibfnamefont {S.}~\bibnamefont {Li}},\ and\ \bibinfo {author} {\bibfnamefont {B.}~\bibnamefont {Liu}},\ }\bibfield  {title} {\bibinfo {title} {Nonlinearity-induced dynamical self-organized twisted-bilayer lattices in bose-einstein condensates},\ }\href {https://arxiv.org/abs/2407.21466} {\bibfield  {journal} {\bibinfo  {journal} {e-print arXiv: 2407.21466}\ } (\bibinfo {year} {2024})}\BibitemShut {NoStop}%
\bibitem [{\citenamefont {Fang}\ \emph {et~al.}(2025)\citenamefont {Fang}, \citenamefont {Gao},\ and\ \citenamefont {Lin}}]{fang2025bifurcations}%
  \BibitemOpen
  \bibfield  {author} {\bibinfo {author} {\bibfnamefont {P.}~\bibnamefont {Fang}}, \bibinfo {author} {\bibfnamefont {C.}~\bibnamefont {Gao}},\ and\ \bibinfo {author} {\bibfnamefont {J.}~\bibnamefont {Lin}},\ }\bibfield  {title} {\bibinfo {title} {Bifurcations and dynamics of nonlinear excitations in twisted-bilayer optical lattices},\ }\href {https://doi.org/10.1016/j.chaos.2025.116314} {\bibfield  {journal} {\bibinfo  {journal} {Chaos, Solitons \& Fractals}\ }\textbf {\bibinfo {volume} {195}},\ \bibinfo {pages} {116314} (\bibinfo {year} {2025})}\BibitemShut {NoStop}%
\bibitem [{\citenamefont {Arkhipova}\ \emph {et~al.}(2023)\citenamefont {Arkhipova}, \citenamefont {Kartashov}, \citenamefont {Ivanov}, \citenamefont {Zhuravitskii}, \citenamefont {Skryabin}, \citenamefont {Dyakonov}, \citenamefont {Kalinkin}, \citenamefont {Kulik}, \citenamefont {Kompanets}, \citenamefont {Chekalin}, \citenamefont {Ye}, \citenamefont {Konotop}, \citenamefont {Torner},\ and\ \citenamefont {Zadkov}}]{23PRL}%
  \BibitemOpen
  \bibfield  {author} {\bibinfo {author} {\bibfnamefont {A.~A.}\ \bibnamefont {Arkhipova}}, \bibinfo {author} {\bibfnamefont {Y.~V.}\ \bibnamefont {Kartashov}}, \bibinfo {author} {\bibfnamefont {S.~K.}\ \bibnamefont {Ivanov}}, \bibinfo {author} {\bibfnamefont {S.~A.}\ \bibnamefont {Zhuravitskii}}, \bibinfo {author} {\bibfnamefont {N.~N.}\ \bibnamefont {Skryabin}}, \bibinfo {author} {\bibfnamefont {I.~V.}\ \bibnamefont {Dyakonov}}, \bibinfo {author} {\bibfnamefont {A.~A.}\ \bibnamefont {Kalinkin}}, \bibinfo {author} {\bibfnamefont {S.~P.}\ \bibnamefont {Kulik}}, \bibinfo {author} {\bibfnamefont {V.~O.}\ \bibnamefont {Kompanets}}, \bibinfo {author} {\bibfnamefont {S.~V.}\ \bibnamefont {Chekalin}}, \bibinfo {author} {\bibfnamefont {F.}~\bibnamefont {Ye}}, \bibinfo {author} {\bibfnamefont {V.~V.}\ \bibnamefont {Konotop}}, \bibinfo {author} {\bibfnamefont {L.}~\bibnamefont {Torner}},\ and\ \bibinfo {author} {\bibfnamefont {V.~N.}\ \bibnamefont {Zadkov}},\ }\bibfield  {title} {\bibinfo {title} {Observation of
  linear and nonlinear light localization at the edges of moir\'e arrays},\ }\href {https://doi.org/10.1103/PhysRevLett.130.083801} {\bibfield  {journal} {\bibinfo  {journal} {Physical Review Letters}\ }\textbf {\bibinfo {volume} {130}},\ \bibinfo {pages} {083801} (\bibinfo {year} {2023})}\BibitemShut {NoStop}%
\bibitem [{\citenamefont {Wang}\ \emph {et~al.}(2024)\citenamefont {Wang}, \citenamefont {Gao}, \citenamefont {Zhang}, \citenamefont {Zhai},\ and\ \citenamefont {Shi}}]{Wang24PRL}%
  \BibitemOpen
  \bibfield  {author} {\bibinfo {author} {\bibfnamefont {C.}~\bibnamefont {Wang}}, \bibinfo {author} {\bibfnamefont {C.}~\bibnamefont {Gao}}, \bibinfo {author} {\bibfnamefont {J.}~\bibnamefont {Zhang}}, \bibinfo {author} {\bibfnamefont {H.}~\bibnamefont {Zhai}},\ and\ \bibinfo {author} {\bibfnamefont {Z.-Y.}\ \bibnamefont {Shi}},\ }\bibfield  {title} {\bibinfo {title} {Three-dimensional moir\'e crystal in ultracold atomic gases},\ }\href {https://doi.org/10.1103/PhysRevLett.133.163401} {\bibfield  {journal} {\bibinfo  {journal} {Physical Review Letters}\ }\textbf {\bibinfo {volume} {133}},\ \bibinfo {pages} {163401} (\bibinfo {year} {2024})}\BibitemShut {NoStop}%
\bibitem [{SM()}]{SM}%
  \BibitemOpen
  \href@noop {} {}\bibinfo {note} {See supplementary material for details.}\BibitemShut {Stop}%
\bibitem [{lin()}]{linear_independence}%
  \BibitemOpen
  \href@noop {} {}\bibinfo {note} {That is to say, the only rational solution to the equation $x\cdot l_2/l_1+ y\cdot l_3/l_1=0$ is the trivial one $x=y=0$.}\BibitemShut {Stop}%
\bibitem [{Kep()}]{Kepler}%
  \BibitemOpen
  \href@noop {} {}\bibinfo {note} {See \href{https://en.wikipedia.org/wiki/Kepler_triangle}{https://en.wikipedia.org/wiki/Kepler\_triangle} for an introduction to the geometric and arithmetic properties of Kepler triangle.}\BibitemShut {Stop}%
\bibitem [{\citenamefont {Huang}\ \emph {et~al.}(2016)\citenamefont {Huang}, \citenamefont {Ye}, \citenamefont {Chen}, \citenamefont {Kartashov}, \citenamefont {Konotop},\ and\ \citenamefont {Torner}}]{huang2016localization}%
  \BibitemOpen
  \bibfield  {author} {\bibinfo {author} {\bibfnamefont {C.}~\bibnamefont {Huang}}, \bibinfo {author} {\bibfnamefont {F.}~\bibnamefont {Ye}}, \bibinfo {author} {\bibfnamefont {X.}~\bibnamefont {Chen}}, \bibinfo {author} {\bibfnamefont {Y.~V.}\ \bibnamefont {Kartashov}}, \bibinfo {author} {\bibfnamefont {V.~V.}\ \bibnamefont {Konotop}},\ and\ \bibinfo {author} {\bibfnamefont {L.}~\bibnamefont {Torner}},\ }\bibfield  {title} {\bibinfo {title} {Localization-delocalization wavepacket transition in pythagorean aperiodic potentials},\ }\href {https://doi.org/10.1038/srep32546} {\bibfield  {journal} {\bibinfo  {journal} {Scientific Reports}\ }\textbf {\bibinfo {volume} {6}},\ \bibinfo {pages} {32546} (\bibinfo {year} {2016})}\BibitemShut {NoStop}%
\bibitem [{\citenamefont {Huang}\ and\ \citenamefont {Liu}(2019)}]{Huang19PRB}%
  \BibitemOpen
  \bibfield  {author} {\bibinfo {author} {\bibfnamefont {B.}~\bibnamefont {Huang}}\ and\ \bibinfo {author} {\bibfnamefont {W.~V.}\ \bibnamefont {Liu}},\ }\bibfield  {title} {\bibinfo {title} {Moir\'e localization in two-dimensional quasiperiodic systems},\ }\href {https://doi.org/10.1103/PhysRevB.100.144202} {\bibfield  {journal} {\bibinfo  {journal} {Physical Review B}\ }\textbf {\bibinfo {volume} {100}},\ \bibinfo {pages} {144202} (\bibinfo {year} {2019})}\BibitemShut {NoStop}%
\bibitem [{\citenamefont {Moon}\ \emph {et~al.}(2019)\citenamefont {Moon}, \citenamefont {Koshino},\ and\ \citenamefont {Son}}]{Moon19PRB}%
  \BibitemOpen
  \bibfield  {author} {\bibinfo {author} {\bibfnamefont {P.}~\bibnamefont {Moon}}, \bibinfo {author} {\bibfnamefont {M.}~\bibnamefont {Koshino}},\ and\ \bibinfo {author} {\bibfnamefont {Y.-W.}\ \bibnamefont {Son}},\ }\bibfield  {title} {\bibinfo {title} {Quasicrystalline electronic states in ${30}^{\ensuremath{\circ}}$ rotated twisted bilayer graphene},\ }\href {https://doi.org/10.1103/PhysRevB.99.165430} {\bibfield  {journal} {\bibinfo  {journal} {Physical Review B}\ }\textbf {\bibinfo {volume} {99}},\ \bibinfo {pages} {165430} (\bibinfo {year} {2019})}\BibitemShut {NoStop}%
\bibitem [{\citenamefont {Park}\ \emph {et~al.}(2019)\citenamefont {Park}, \citenamefont {Kim},\ and\ \citenamefont {Lee}}]{Park19PRB}%
  \BibitemOpen
  \bibfield  {author} {\bibinfo {author} {\bibfnamefont {M.~J.}\ \bibnamefont {Park}}, \bibinfo {author} {\bibfnamefont {H.~S.}\ \bibnamefont {Kim}},\ and\ \bibinfo {author} {\bibfnamefont {S.}~\bibnamefont {Lee}},\ }\bibfield  {title} {\bibinfo {title} {Emergent localization in dodecagonal bilayer quasicrystals},\ }\href {https://doi.org/10.1103/PhysRevB.99.245401} {\bibfield  {journal} {\bibinfo  {journal} {Physical Review B}\ }\textbf {\bibinfo {volume} {99}},\ \bibinfo {pages} {245401} (\bibinfo {year} {2019})}\BibitemShut {NoStop}%
\bibitem [{\citenamefont {Tang}\ \emph {et~al.}(2021)\citenamefont {Tang}, \citenamefont {Du}, \citenamefont {Carr}, \citenamefont {DeVault}, \citenamefont {Mello},\ and\ \citenamefont {Mazur}}]{tang2021modeling}%
  \BibitemOpen
  \bibfield  {author} {\bibinfo {author} {\bibfnamefont {H.}~\bibnamefont {Tang}}, \bibinfo {author} {\bibfnamefont {F.}~\bibnamefont {Du}}, \bibinfo {author} {\bibfnamefont {S.}~\bibnamefont {Carr}}, \bibinfo {author} {\bibfnamefont {C.}~\bibnamefont {DeVault}}, \bibinfo {author} {\bibfnamefont {O.}~\bibnamefont {Mello}},\ and\ \bibinfo {author} {\bibfnamefont {E.}~\bibnamefont {Mazur}},\ }\bibfield  {title} {\bibinfo {title} {Modeling the optical properties of twisted bilayer photonic crystals},\ }\href {https://doi.org/s41377-021-00601-x} {\bibfield  {journal} {\bibinfo  {journal} {Light: Science \& Applications}\ }\textbf {\bibinfo {volume} {10}},\ \bibinfo {pages} {157} (\bibinfo {year} {2021})}\BibitemShut {NoStop}%
\bibitem [{\citenamefont {Gonçalves}\ \emph {et~al.}(2021)\citenamefont {Gonçalves}, \citenamefont {Olyaei}, \citenamefont {Amorim}, \citenamefont {Mondaini}, \citenamefont {Ribeiro},\ and\ \citenamefont {Castro}}]{Gonçalves_2022}%
  \BibitemOpen
  \bibfield  {author} {\bibinfo {author} {\bibfnamefont {M.}~\bibnamefont {Gonçalves}}, \bibinfo {author} {\bibfnamefont {H.~Z.}\ \bibnamefont {Olyaei}}, \bibinfo {author} {\bibfnamefont {B.}~\bibnamefont {Amorim}}, \bibinfo {author} {\bibfnamefont {R.}~\bibnamefont {Mondaini}}, \bibinfo {author} {\bibfnamefont {P.}~\bibnamefont {Ribeiro}},\ and\ \bibinfo {author} {\bibfnamefont {E.~V.}\ \bibnamefont {Castro}},\ }\bibfield  {title} {\bibinfo {title} {Incommensurability-induced sub-ballistic narrow-band-states in twisted bilayer graphene},\ }\href {https://doi.org/10.1088/2053-1583/ac3259} {\bibfield  {journal} {\bibinfo  {journal} {2D Materials}\ }\textbf {\bibinfo {volume} {9}},\ \bibinfo {pages} {011001} (\bibinfo {year} {2021})}\BibitemShut {NoStop}%
\bibitem [{\citenamefont {Gao}\ \emph {et~al.}(2023)\citenamefont {Gao}, \citenamefont {Xu}, \citenamefont {Yang},\ and\ \citenamefont {Ye}}]{Gao23PRA}%
  \BibitemOpen
  \bibfield  {author} {\bibinfo {author} {\bibfnamefont {Z.}~\bibnamefont {Gao}}, \bibinfo {author} {\bibfnamefont {Z.}~\bibnamefont {Xu}}, \bibinfo {author} {\bibfnamefont {Z.}~\bibnamefont {Yang}},\ and\ \bibinfo {author} {\bibfnamefont {F.}~\bibnamefont {Ye}},\ }\bibfield  {title} {\bibinfo {title} {Pythagoras superposition principle for localized eigenstates of two-dimensional moir\'e lattices},\ }\href {https://doi.org/10.1103/PhysRevA.108.013513} {\bibfield  {journal} {\bibinfo  {journal} {Physical Review A}\ }\textbf {\bibinfo {volume} {108}},\ \bibinfo {pages} {013513} (\bibinfo {year} {2023})}\BibitemShut {NoStop}%
\bibitem [{\citenamefont {Paul}\ \emph {et~al.}(2024)\citenamefont {Paul}, \citenamefont {Crowley},\ and\ \citenamefont {Fu}}]{24PRB}%
  \BibitemOpen
  \bibfield  {author} {\bibinfo {author} {\bibfnamefont {N.}~\bibnamefont {Paul}}, \bibinfo {author} {\bibfnamefont {P.~J.~D.}\ \bibnamefont {Crowley}},\ and\ \bibinfo {author} {\bibfnamefont {L.}~\bibnamefont {Fu}},\ }\bibfield  {title} {\bibinfo {title} {Directional localization from a magnetic field in moir\'e systems},\ }\href {https://doi.org/10.1103/PhysRevLett.132.246402} {\bibfield  {journal} {\bibinfo  {journal} {Physical Review Letters}\ }\textbf {\bibinfo {volume} {132}},\ \bibinfo {pages} {246402} (\bibinfo {year} {2024})}\BibitemShut {NoStop}%
\bibitem [{\citenamefont {Madro{\~n}ero}\ \emph {et~al.}(2024)\citenamefont {Madro{\~n}ero}, \citenamefont {Castro},\ and\ \citenamefont {Paredes}}]{madroñero2024}%
  \BibitemOpen
  \bibfield  {author} {\bibinfo {author} {\bibfnamefont {C.}~\bibnamefont {Madro{\~n}ero}}, \bibinfo {author} {\bibfnamefont {G.~A.~D.}\ \bibnamefont {Castro}},\ and\ \bibinfo {author} {\bibfnamefont {R.}~\bibnamefont {Paredes}},\ }\bibfield  {title} {\bibinfo {title} {Localized and extended phases in square moir\'e patterns},\ }\href {https://arxiv.org/abs/2405.00811} {\bibfield  {journal} {\bibinfo  {journal} {e-print arXiv: 2405.00811}\ } (\bibinfo {year} {2024})}\BibitemShut {NoStop}%
\end{thebibliography}%

\end{document}